\documentclass[runningheads]{llncs}
\usepackage{epsfig}
\usepackage{amsmath}
\usepackage{amssymb}

\newcommand{\denselist}{\itemsep 0pt\parsep=1pt\partopsep 0pt}
\newcommand{\mm}[1] {\ifmmode{#1}\else{\mbox{\(#1\)}}\fi}

\newcommand{\eop}{\hbox{}\nobreak\hfill\quad\usebox{\smallProofsym}\bigskip}  %
\newsavebox{\smallProofsym}                            
\savebox{\smallProofsym}                               %
{                                                      %
\begin{picture}(6,6)                                   %
\put(0,0){\framebox(6,6){}}                            %
\put(0,2){\framebox(4,4){}}                            %
\end{picture}                                          %
}                                                      %

\newcommand{\Rspace}        {\mm{{\mathbb R}}}
\newcommand{\Fset}          {\mm{\cal F}}

\newcommand{\Eps}           {\mm{\varepsilon}}

\newcommand{\norm}[1]       {\mm{\|{#1}\|}}
\newcommand{\dist}[2]       {\mm{\|{#1}-{#2}\|}}
\newcommand{\wdist}[1]      {\mm{f_{#1}}}

\newcommand{\env}[1]        {\mm{\rm env\,}{#1}}
\newcommand{\diff}          {\mm{\rm d}}

\newif\ifpdf\ifx\pdfoutput\undefined\pdffalse\else\pdfoutput=1\pdftrue\fi
\newcommand{\pdfgraphics}{\ifpdf\DeclareGraphicsExtensions{.pdf,.jpg}\else\fi}
 
\mainmatter         

\title{Relaxed Scheduling in\\ Dynamic Skin Triangulation
	\thanks{Research of the two authors is supported by
		NSF under grant CCR-00-86013.}}

\author{
   Herbert Edelsbrunner \inst{1}
    \and
   Alper \"{U}ng\"{o}r \inst{2}
}

\institute{
    Department of Computer Science,
    Duke University, Durham, NC 27708, and  Raindrop Geomagic,
    Research Triangle Park, NC 27709,\\
    \email{edels@cs.duke.edu} \vspace{.07in}
  \and
    Department of Computer Science,
    Duke University, Durham, NC 27708,\\
    \email{ungor@cs.duke.edu}
}

\begin{document}
\pdfgraphics

\maketitle

\begin{abstract}
{\rm
  We introduce relaxed scheduling as a paradigm for
  mesh maintenance and demonstrate its applicability to
  triangulating a skin surface in $\Rspace^3$.
}
\end{abstract}

\vspace{0.1in}

\section{Introduction}
\label{sec1}

In this paper, we describe a relaxed scheduling paradigm for operations
that maintain the mesh of a deforming surface.
We prove the correctness of this paradigm for skin surfaces.

\begin{figure}[hbt]
\vspace*{0.1in}
 \centering
  \centerline{
     \includegraphics[height=1.6in]{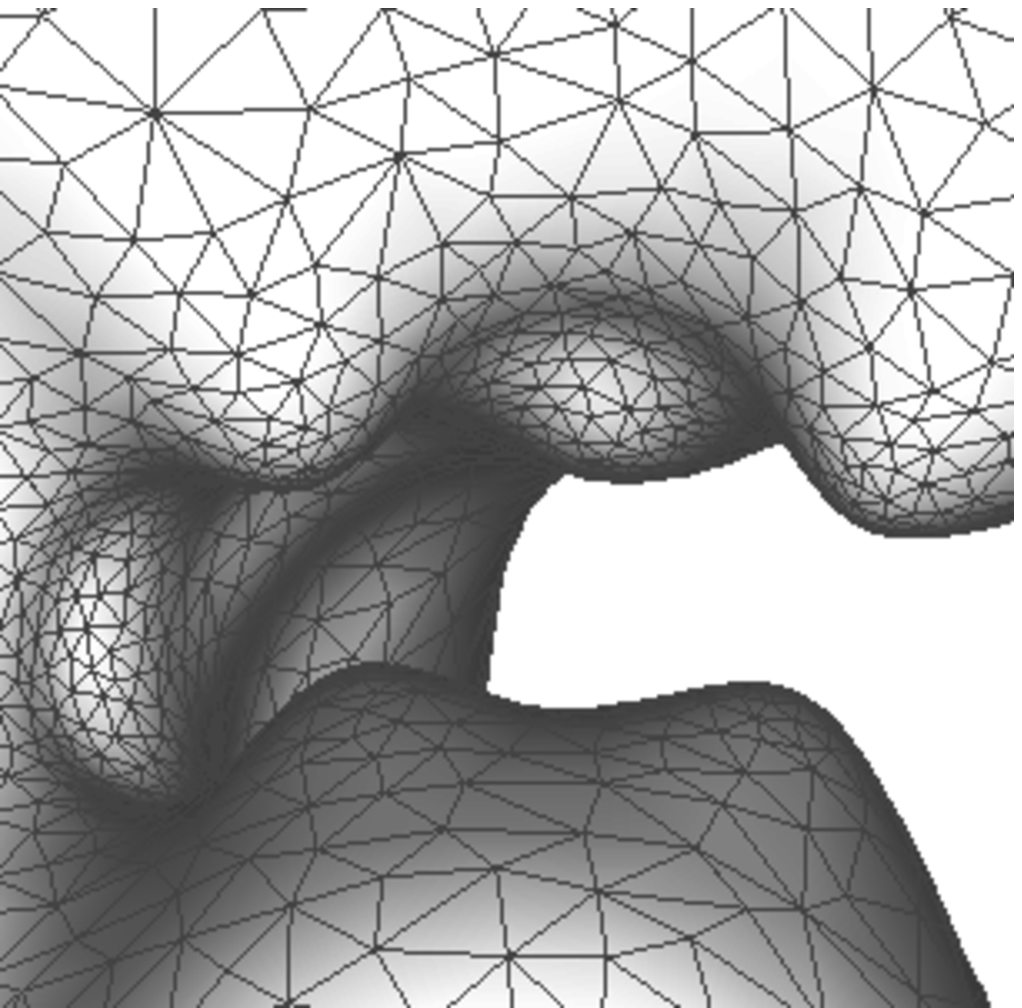} \hspace{.15in}
     \includegraphics[height=1.6in]{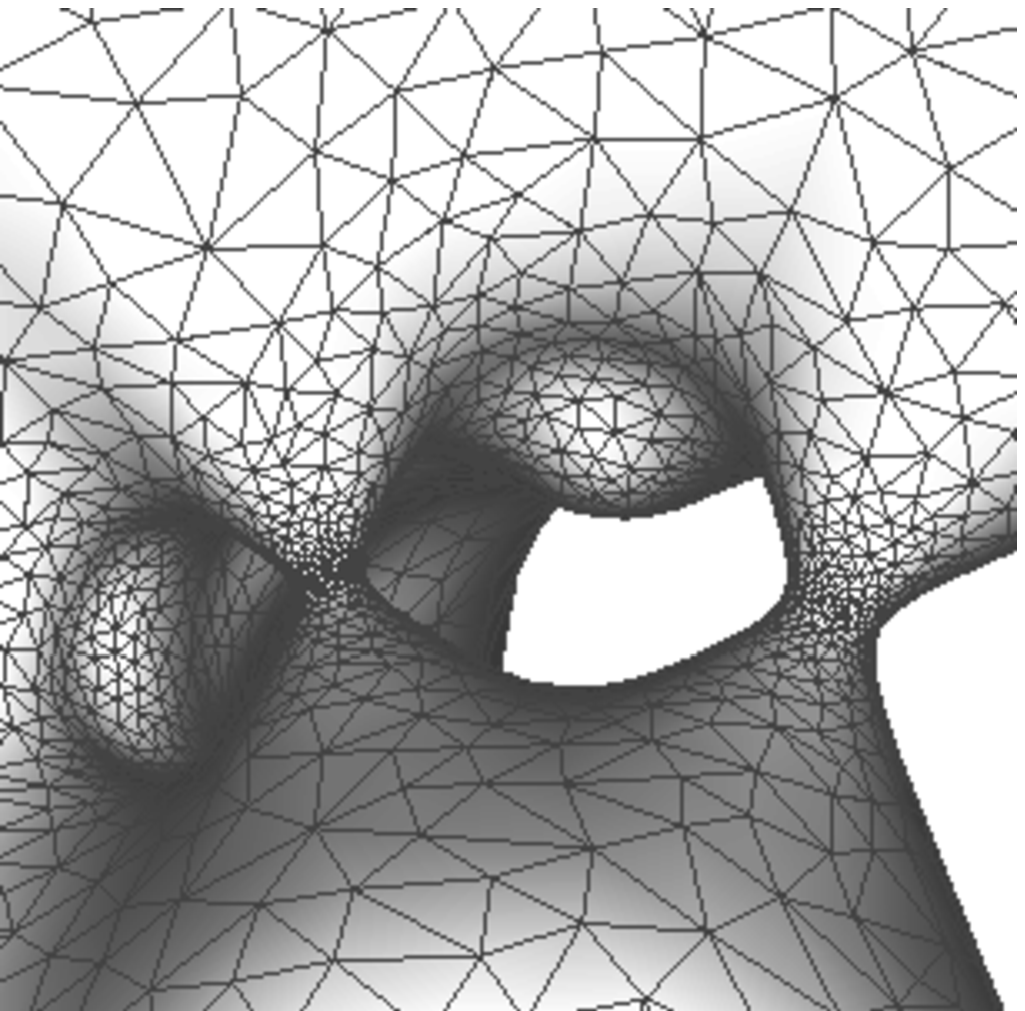} }
 \caption{The mesh is maintained as the surface on the left grows
          into that on the right}
 \label{fig:deformation}
\end{figure}

\paragraph{Background.}
In 1999, Edelsbrunner \cite{Ede99} showed how a finite collection of spheres
or weighted points can be used to construct a $C^1$-continuous surface
in $\Rspace^3$.
It is referred to as the \emph{skin} or the \emph{skin surface} of the collection.
If the spheres represent the atoms of a molecule then the appearance of that
surface is similar to the molecular surface used in
structural biology \cite{Con83,LeRi71}.
The two differ in a number of details, one being that the former uses
hyperboloids to blend between sphere patches while the latter uses tori.
The skin surface is not $C^2$-continuous,
but its maximum normal curvature, $\kappa$, is continuous.
This property is exploited by Cheng {\em et al.}\ \cite{CDES01}, who describe
an algorithm that constructs a triangular mesh representing the skin surface.
In this mesh, the sizes of edges and triangles are inversely proportional
to the maximum normal curvature.
The main idea of the algorithm is to maintain the mesh while gradually growing
the skin surface to the desired shape,
as illustrated in Figure \ref{fig:deformation}.
The algorithm thus reduces the construction to a sequence of restructuring operations.
There are \emph{edge flips}, which maintain the mesh as the restricted Delaunay
triangulation of its vertices,
\emph{edge contractions} and \emph{vertex insertions},
which maintain a sampling whose local density is proportional to the maximum normal curvature,
and \emph{metamorphoses}, which adjust the mesh connectivity to
reflect changes in the surface topology.
Some of these operations are easier to schedule than others,
and the most difficult ones are the edge contractions and vertex insertions.
They depend on how the sampled points move with the surface as it deforms.
The quality of the mesh is guaranteed by maintaining size constraints
for all edges and triangles.
When an edge gets too short we contract it, and when a triangle gets too large
we insert a point near its circumcenter.
Both events can be recognized by finding roots of fairly involved functions.
Scheduling edge contractions and vertex insertions thus becomes a bottleneck,
both in terms of the robustness and the running time of the algorithm.

\paragraph{Result.}
In this paper, we study how fast edges and triangles vary their size,
and we use that knowledge to schedule these elements in a relaxed fashion.
In other words, we do not determine when exactly an element violates
its size constraint, but we catch it before the violation happens.
Of course, the danger is now that we either update perfectly well-shaped
elements or we waste time by checking elements unnecessarily often.
To avoid the former, we introduce intervals or gray zones in which
the shapes of the elements are neither good nor unacceptably bad.
To avoid unnecessarily frequent checking, we prove lower bounds on how long an
element stays in the gray zone before its shape becomes unacceptably bad.
These bounds are different for edges and for triangles.
Consider first an edge $uv$.
Let $R = \dist{u}{v} / 2$ be its half-length and
$\varrho = 1 / \max \{\kappa(u), \kappa(v)\}$ the smaller radius of curvature
at its endpoints.
We use judiciously chosen constants $C$, $Q_0$ and $Q_1$ and call the edge
$$
  \left . \begin{array}{l}
    \mbox{\it acceptable}  \\  \mbox{\it borderline}  \\  \mbox{\it unacceptable}
  \end{array} \right \}
  \mbox{\rm ~~~~if~~~~}
  \left \{ \begin{array}{l}
    C/Q_0   <       R/\varrho             , \\
    C/Q_1   <       R/\varrho  \leq  C/Q_0  , \\
       ~~~~~~~~~~~\,  R/\varrho  \leq  C/Q_1 .
  \end{array} \right .
$$
The middle interval is what we called the gray zone above.
Assuming $uv$ is acceptable, we prove it will not become unacceptable within
a time interval of duration $\Delta t = (2 \theta - \theta^2) \varrho^2$, where
\begin{eqnarray*}
  \theta ~=~ \frac{R Q_1 - C \varrho}{R Q_1 + C \varrho} .
\end{eqnarray*}
In the worst case, $R$ is barely larger than $C \varrho / Q_0$, so we have
$\theta > (Q_1 - Q_0) / (Q_1 + Q_0)$ as a worst case bound.
We will see that $C = 0.06$, $Q_0 = 1.6$ and $Q_1 = 2.3$ are feasible choices for
the constants, and that for these we get $\theta > 0.179 \ldots$ and
$\Delta t / \varrho^2 > 0.326 \ldots$.
Consider next a triangle $uvw$.
Let $R$ be the radius of its circumcircle, and
$\varrho = 1 / \max \{ \kappa(u), \kappa(v), \kappa(w) \}$ the smallest
radius of curvature at its vertices.
We call $uvw$
$$
  \left . \begin{array}{l}
    \mbox{\it acceptable}  \\  \mbox{\it borderline}  \\  \mbox{\it unacceptable}
  \end{array} \right \}
  \mbox{\rm ~~~~if~~~~}
  \left \{ \begin{array}{l}
    ~~~~~~~~~~  R/\varrho    <   C Q_0 ,  \\
    C Q_0    \leq   R/\varrho    <   CQ_1 ,   \\
    C Q_1    \leq   R/\varrho .
  \end{array} \right .
$$
Assuming $uvw$ is acceptable, we prove it will not become unacceptable within
a time interval of duration $\Delta t = (2 \theta - \theta^2) \varrho^2$, where
\begin{eqnarray*}
  \theta  ~=~ 1 - \sqrt[4]{R / (C Q_1 \varrho)} .
\end{eqnarray*} 
In the worst case, $R$ is barely smaller than $C Q_0 \varrho$, so we have
$\theta > 1 - \sqrt[4]{Q_0 / Q_1}$. 
For the above values of $C$, $Q_0$ and $Q_1$, this gives
$\theta > 0.086\ldots$ and $\Delta t / \varrho^2 > 0.165\ldots$.
It seems that triangles can get out of shape about twice as fast as edges,
but we do not know whether this is really the case because our bounds are not tight.

\paragraph{Outline.}
Section \ref{sec2} reviews skin surfaces and the dynamic triangulation algorithm.
Section \ref{sec3} introduces relaxed scheduling as a paradigm to keep track of
moving or deforming data.
Section \ref{sec4} analyzes the local distortion within the mesh
and derives the formulas needed for the relaxed scheduling paradigm.
Section \ref{sec5} concludes the paper.

\section{Preliminaries}
\label{sec2}

In this section, we introduce the necessary background from \cite{Ede99},
where skin surfaces were originally defined, and from \cite{CDES01},
where the meshing algorithm for deforming skin surfaces was described.

\begin{figure}[hbt]
  \vspace*{0.1in}
  \centering
  \centerline{\includegraphics[height=2.0in]{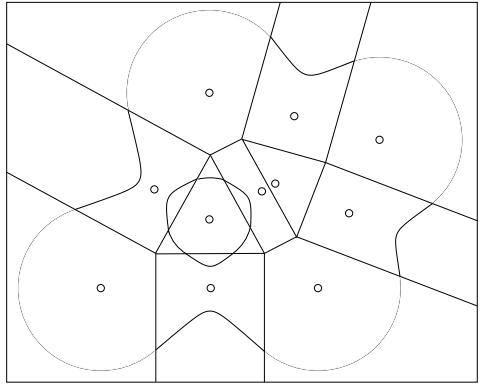}}
  \caption{The mixed complex decomposes the skin curve and the area it bounds}
  \label{fig:mixed}
\end{figure}

\paragraph{Skin surfaces.}
We write $S_i = (z_i, r_i)$ for the sphere with center $z_i \in \Rspace^3$
and radius $r_i$ and think of it as the zero-set of the weighted square
distance function $\wdist{i}: \Rspace^3 \rightarrow \Rspace$ defined by
$\wdist{i} (x) = \dist{x}{z_i}^2 - r_i^2$.
The square radius is a real number and the radius is either a
non-negative real or a non-negative multiple of the imaginary unit.
We know how to add functions and how to multiply them by scalars.
For example, if we have a finite collection of spheres $S_i$ and
scalars $\sum \gamma_i = 1$ then
$\sum \gamma_i \wdist{i}$ is again a weighted square distance function,
and we denote by $S = \sum \gamma_i S_i$ the sphere that defines it.
The \emph{convex hull} of the $S_i$ is the set of such spheres obtained
using only non-negative scalars:
\begin{eqnarray*}
  \Fset ~=~   \left\{ \sum \gamma_i S_i  \mid
                      \sum \gamma_i = 1 \mbox{\rm ~and~}
                      \gamma_i \geq 0, \forall i \right\} .
\end{eqnarray*}
We also shrink spheres and write $\sqrt{S} = (z, r/\sqrt{2})$,
which is the zero-set of $2 \wdist{} - \wdist{} (z)$.
The \emph{skin surface} defined by the $S_i$
is then the envelope of the spheres in the convex hull, all scaled down by
a factor $1/\sqrt{2}$, and we write this as $F  =  \env{\sqrt{\Fset}}$.
Equivalently, it is the zero-set of the point-wise minimum over all functions
$2 \wdist{} - \wdist{} (z)$, over all $S \in \Fset$,
where $\wdist{}$ is the weighted square distance function defined by $S$.
At first glance, this might seem like an unwieldy surface, but we can
completely describe it as a collection of quadratic patches obtained by
decomposing the surface with what we call the mixed complex.
Its cells are Minkowski sums of Voronoi vertices, edges, polygons and
polyhedra with their dually corresponding
Delaunay tetrahedra, triangles, edges and vertices,
all scaled down by a factor $1/2$.
Instead of formally describing this construction, we illustrate it with a
two-dimensional example in Figure \ref{fig:mixed}.
Depending on the dimension of the contributing Delaunay simplex,
we have four types of mixed cells.
Because of symmetry, we have only two types of surface patches, namely pieces
of spheres and of hyperboloids of revolution, which we frequently put in
Standard Form:
\begin{eqnarray}
  \xi_1^2 + \xi_2^2 + \xi_3^2 ~&=&~ R^2 ,  \label{eqn:standardsphere} \\ 
  \xi_1^2 + \xi_2^2 - \xi_3^2 ~&=&~ \pm R^2 , \label{eqn:standardhyperboloid}
\end{eqnarray}
where the plus sign gives the one-sheeted hyperboloid and the minus sign
gives the two-sheeted hyperboloid.

\paragraph{Meshing.}
The meshing algorithm triangulates the skin surface using edges and triangles
whose sizes adapt to the local curvature. Let us be more specific.
At any point $x \in F$, let $\kappa (x)$ be the maximum normal curvature at $x$.
In contrast to other notions of curvature, $\kappa$ is continuous over
the skin surface and thus amenable to controlling the local size of the mesh.
Call $\varrho (x) = 1/ \kappa (x)$ the \emph{local length scale} at $x$.
The vertices of the mesh are points on the surface.
For an edge $uv$, let $R_{uv} = \dist{u}{v} /2$ be half its length,
and for a triangle $uvw$, let $R_{uvw}$ be the radius of its circumcircle.
The algorithm obeys the Lower and Upper Size Bounds that require
edges not be too short and triangles not be too large:
\begin{description}\denselist
  \item[{\rm [L]}] $R_{uv} / \varrho_{uv} ~>~ C/Q$ for every
                   edge $uv$, and
  \item[{\rm [U]}] $R_{uvw} / \varrho_{uvw} ~<~ CQ$ for every triangle $uvw$,
\end{description}
where $\varrho_{uv}$ is the larger of $\varrho (u)$ and $\varrho (v)$,
$\varrho_{uvw}$ is the minimum of $\varrho (u)$, $\varrho (v)$ and $\varrho (w)$,
and $C$ and $Q$ are judiciously chosen positive constants.

The particular algorithm we consider in this paper is dynamic,
in the sense that it maintains the mesh while the surface deforms.
We can use this algorithm to construct a mesh by starting with the empty surface
and growing it into the desired shape.
This is precisely the scenario in which our results apply.
To model the growth process, we use a time parameter and
let $S_i (t) = (z_i, \sqrt{r_i^2 + t})$ be the $i$-th sphere
at time $t \in \Rspace$.
We start at $t = - \infty$, at which time all radii are imaginary
and the surface is empty, and we end at $t = 0$,
at which time the surface has the desired shape.
This particular growth model is amenable to efficient computation because
it does not affect the mixed complex, which stays the same at all times.
Each patch of the surface sweeps out its mixed cell.
At any moment, we have a collection of points sampled on the surface,
and the mesh is the restricted Delaunay triangulation of these points,
as defined in \cite{Che93,EdSh97}.
Given the surface and the points, this triangulation is unique.
As the surface deforms, we move the points with it and update the mesh
as required.
From global and less frequent to local and more frequent these operations are:
\begin{enumerate}\denselist
 \item topology changes that affect the local and global connectivity
       of the surface and the mesh,
 \item edge contractions and vertex insertions that locally remove or add
       points to coarsen or refine the mesh, and
 \item edge flips that locally adjust the mesh without affecting the
       point distribution or the surface topology.
\end{enumerate}
For the particular growth model introduced above, the topology changes
are easily predicted using the filtration of alpha complexes as described
in \cite{EdMu94}.
To predict where and when we need to coarsen or refine the mesh is more
difficult and depends on how the points move to follow
the deforming surface.
This is the topic of this paper and will be discussed in detail in
the subsequent sections.
Finally, edge flips are relatively robust operations, which can be
performed in a lazy manner, without any sophisticated scheduling mechanism.

\begin{figure}[hbt]
 \vspace*{0.1in}
 \centering
 \centerline{\includegraphics[height=1.9in]{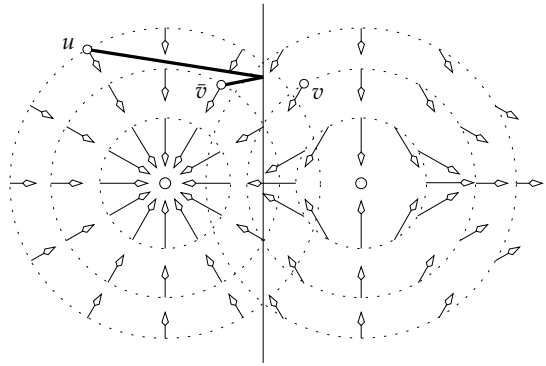}}
 \caption{Velocity vectors of a shrinking circle on the left and
          of a hyperbola on the right.
          The right portion of the edge $uv$ is reflected across
          face shared by the two mixed cells}
 \label{fig:reflection}
\end{figure}

\paragraph{Point motion.}
To describe the motion of the points sampled on the skin surface,
it is convenient to consider the trajectory of the surface over time.
Note that the $i$-th sphere at time $t$ is $S_i (t) = f_i^{-1} (t)$.
Similarly, the convex combination defined by coefficients $\gamma_i$
at time $t$ is $S(t) = f^{-1} (t)$, where $f = \sum \gamma_i f_i$.
We can represent the skin surface in the same manner by introducing
the function $g: \Rspace^3 \rightarrow \Rspace$ defined as the point-wise
minimum of the functions representing the shrunken spheres.
More formally,
$g(x) = \min \{ 2 f(x) - f(z) \}$, where the minimum is taken over all
spheres $S \in \Fset$ and $z$ is the center of $S$.
The skin surface at time $t$ is then $F(t) = g^{-1} (t)$,
so it is appropriate to call the graph of $g$ the \emph{trajectory} of
the skin surface.
We see that growing the surface in time is equivalent to sweeping out its
trajectory with a three-dimensional space that moves through time.
It is natural to let the points sampled on $F(t)$ move normal to the surface.
For a point $x = [\xi_1, \xi_2, \xi_3]^T$ on a sphere or hyperboloid
in Standard Form $\xi_1^2 + \xi_2^2 \pm \xi_3^2 = \pm R^2$,
the gradient is $\nabla g_x = 2  [\xi_1, \xi_2, \pm \xi_3]^T$.
The point $x$ moves in the direction of the gradient with a speed that is
inversely proportional to the length.
In other words, the velocity vector at a point $x$ is
\begin{eqnarray*}
  \dot{x} ~=~  \frac{\diff x}{\diff t}
          ~=~  \frac{\nabla g_x}{\norm{\nabla g_x}^2}
          ~=~  \frac{\nabla g_x}{4 \norm{x}^2} .
\end{eqnarray*}
The speed of $x$ is therefore $\norm{\dot{x}} = 1 / (2 \norm{x})$.
The implementation of the relaxed scheduling paradigm crucially depends on the
properties of this motion.
We use the remainder of this section to describe a symmetry property of the
velocity vectors that is instrumental in the analysis of the motion.
Consider two mixed cells that share a common face.
The Standard Forms of the two corresponding surface patches differ by
a single sign, and so do the gradients.
If we reflect points in one cell across the plane of the common face into
the other cell then we preserve the velocity vector, as illustrated in
Figure \ref{fig:reflection}.
We use this observation about adjacent mixed cells to relate the
velocity vectors of points in possibly non-adjacent cells.
Consider points $u$ and $v$ and let $x_1, x_2, \ldots, x_k$
be the intersection points with faces of mixed cells encountered as we
travel along the edge from $u$ to $v$.
Starting at $i = k$, we work backward and reflect the portion of
the edge beyond $x_i$ across the face that contains $x_i$.
In the general case, this portion is a polygonal path that leads from $x_i$
to the possibly multiply reflected image $\overline{v}$ of $v$.
After $k$ reflections we have a polygonal path from $u$ to the final $\overline{v}$.
The length of the path is equal to the length of the initial edge,
and hence $\dist{u}{\overline{v}} \leq \dist{u}{v}$.
We note that $\overline{v}$ does not necessarily lie in the mixed cell of $u$,
but its velocity vector --- which is the same as that of $v$ --- is consistent
with the family of spheres or hyperboloids that sweeps out that mixed cell.
In other words, the motion of $u$ and $\overline{v}$ is determined by
the same quadratic function.

\section{Relaxed Scheduling}
\label{sec3}

In this section, we introduce relaxed scheduling as a paradigm for maintaining
moving or deforming data.
It is designed to cope with situations in which the precise moment for an
update is either not known or too expensive to compute.

\begin{figure}[hbt]
  \vspace*{0.1in}
  \centering
  \centerline{\includegraphics[height=2.3in]{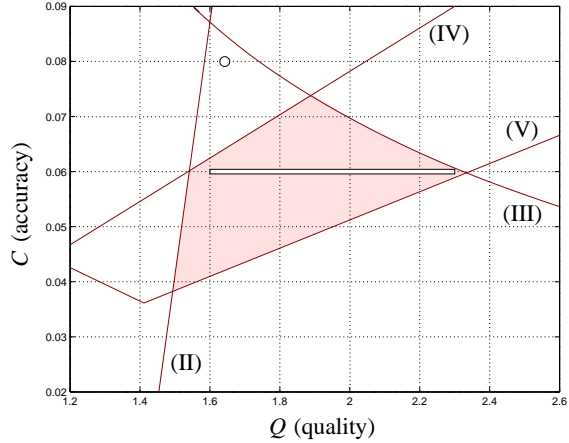}}
  \caption{The shaded feasible region of parameter pairs $(C, Q)$
           for $\Eps = \Eps_0$, $h=0.993$, $\ell=6$ and $m=80$.
           For $C = 0.06$ this region contains the interval $Q \in [1.6, 2.3]$.
           The bounding curves are labeled by the corresponding constraints.
           Redundant constraints are not shown}
  \label{fig:chart}
\end{figure}

\paragraph{Correctness constraints.}
In the context of maintaining the triangle mesh of a skin surface,
we use relaxed scheduling to determine when to contract an edge and
when to insert a new vertex.
Since determining when the size of an edge or triangle stops to be acceptable
is expensive, we introduce a gray zone between acceptability and unacceptability
and update an element when we catch it inside that gray zone.
That this course of action is even conceivable is based on the correctness
proof of the dynamic skin triangulation algorithm for a range of its
controlling parameters.
The first three conditions defining that range refer to $\Eps$, $C$ and $Q$.
We have seen the latter two before in the formulation of the two
Size Bounds [L] and [U]:
$C$ controls how well the mesh approximates the surface, and $Q$ controls the
quality of the mesh.
Both are related to $\Eps$, which quantifies the sampling density.
\begin{description}
 \item[{\rm (I)}]   We require $0 < \Eps \leq \Eps_0$,
                    where $\Eps_0 = 0.279\ldots$ is a root of
                    $2 \cos ( \arcsin \frac{2 \Eps}{1 - \Eps} + \arcsin \Eps )
                          - \frac{2 \Eps}{1 - \Eps} = 0$.
 \item[{\rm (II)}]  $Q^2 - 4CQ > 2$.
 \item[{\rm (III)}] $\frac{\delta^2}{1+\delta^2} - \frac{\delta^2}{4} > C^2 Q^2$,
                    where $\delta = \Eps - \frac{2C(\Eps+1)}{Q+2C}$.
\end{description}
It is computationally efficient to select the loosest possible bound
for the sampling density: $\Eps = \Eps_0$.
Then we get $\delta = 0.166 \ldots$ and, as noted in \cite{CDES01},
we may choose $C = 0.08$ and $Q = 1.65$ to satisfy Conditions (I) to (III).
Alternatively, we may lower $C$ to $0.06$ and are then free to pick $Q$ anywhere
inside the interval from $1.6$ to $2.3$.
The two choices of parameters are marked by a hollow dot and a white bar
in Figure \ref{fig:chart}.
The last two conditions refer to $h$, $\ell$ and $m$.
All three parameters control how metamorphoses that add or remove a handle
are implemented.
Since the curvature blows up at the point and time of a topology change,
we use a special and relatively coarse sampling inside spherical neighborhoods of
such points.
Assuming a unit radius of such neighborhoods, we turn the special sampling
strategy on and off when the skin surface enters and leaves the smaller spherical
neighborhood of radius $h < 1.0$.
If the skin enters as a two-sheeted hyperboloid we triangulate it using two
$\ell$-sided pyramids inside the unit sphere neighborhood.
If it enters as a one-sheeted hyperboloid we triangulate it as an $m$-sided
drum with a waist.
\begin{figure}[hbt]
 \vspace*{0.1in}
 \centering
 \centerline{\includegraphics[height=1.1in]{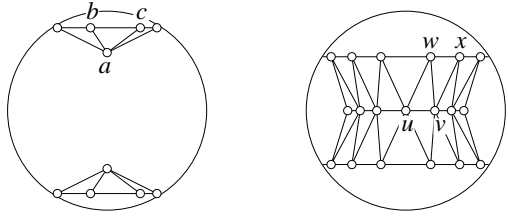}}
 \caption{The triangulation of a two-sheeted and a one-sheeted hyperboloid 
          inside a unit neighborhood sphere around their apices}
 \label{fig:DV}
\end{figure}
The conditions are stated in terms of the edges $ab$, $bc$ and $wx$ and the
triangles $abc$ and $vwx$, as defined in Figure \ref{fig:DV}.
Their sizes can all be expressed in terms of $h$, $\ell$ and $m$,
and we refer to \cite[Section 10]{CDES01} for the formulas.
\begin{description}\denselist
 \item[{\rm (IV)}]  $R_{ab}, R_{bc}, R_{wx} > C/Q$.
 \item[{\rm (V)}]   $R_{abc}, R_{vwx} < \min \{ Q, 2/Q \} Ch$.
\end{description}

\paragraph{Quality buffer.}
The key technical insight about the dynamic skin triangulation algorithm is that
we can find constants $\Eps$, $C$, $h$, $\ell$, $m$ and $Q_0 < Q_1$
such that Conditions (I) to (V) are satisfied for all $Q \in [Q_0, Q_1]$.
This is illustrated in Figure \ref{fig:chart},
which shows the feasible region of points $(C, Q)$ assuming fixed values
for $\Eps$, $h$, $\ell$ and $m$.
Instead of fixing $Q$ and contracting an edge when its size-scale ratio reaches
$C/Q$, we suggest to contract the edge any time its ratio is in the interval
$(C/Q_1, C/Q_0]$.
After the ratio enters this interval at $C/Q_0$ it can either leave again at $C/Q_0$
or it can get contracted, but it is not allowed to reach $C/Q_1$.
Vertex insertions are treated symmetrically.
Specifically, a triangle is removed by adding a vertex near its circumcenter,
and this can happen at any moment its size-scale ratio is in $[CQ_0, CQ_1)$.
The ratio can enter and leave the interval at $CQ_0$, but it is not allowed
to reach $CQ_1$.
We call $(C/Q_1, C/Q_0]$ and $[CQ_0, CQ_1)$ the \emph{lower} and
\emph{upper size buffers}.
The quality of the mesh is guaranteed because all edges and triangles satisfy
the two Size Bounds [L] and [U] for $Q = Q_1$.
Symmetrically, the correctness of the triangulation is guaranteed because
edge contractions and vertex insertions are executed only if the same bounds
are violated for $Q = Q_0$.

\begin{figure}[hbt]
  \vspace*{0.1in}
  \centering
  \centerline{\includegraphics[height=2.3in]{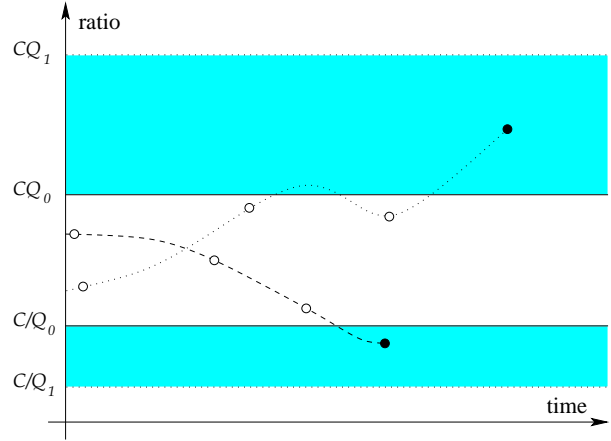}}
  \caption{The two buffers are shaded and the two curves are possible
           developments of size-scale ratios for an edge (dashed)
           and a triangle (dotted).
           The dots indicate moments at which the elements are tested
           and finally removed}
  \label{fig:buffers}
\end{figure}

\paragraph{Early warning.}
Recall that an edge is borderline iff its size-scale ratio is contained in the
lower size buffer, and it becomes unacceptable at the moment it reaches $C/Q_1$.
Similarly, a triangle is borderline iff its size-scale ratio is contained in the
upper size buffer, and it becomes unacceptable at the moment it reaches $CQ_1$.
The relaxed scheduling paradigm depends on an early warning algorithm that reports
an element before it becomes unacceptable.
That algorithm might err and produce false positives, but it may not let any
element slip by and become unacceptable.
False positives cost time but do not cause any harm, while unacceptable elements
compromise the correctness of the meshing algorithm.
In Figure \ref{fig:buffers}, false positives are marked by hollow dots and deletions
are marked by filled black dots.
All false positive tests of edges are represented by dots above the lower
size buffer.
To get a correct early warning algorithm we just need to test each edge
often enough so that its size-scale ratio cannot cross the entire lower size buffer
between two contiguous tests.
The symmetric rule applies to triangles.
Bounds on the amount of time it takes to cross the size buffers will be
given in Section \ref{sec4}.

Note that we have selected the parameters to obtain a fairly long
interval $[Q_0, Q_1]$.
It is not clear whether or not this is a good idea or whether a shorter interval
would lead to a more efficient algorithm.
An argument \emph{for} a long interval is that the implied large size buffers
let us get by with less frequent and therefore fewer tests.
An argument \emph{against} a long interval is that large size buffers are more
likely to cause the deletion of elements that are on their way to better
health but did not recover fast enough and get caught before they could
leave the buffers.
It might be useful to optimize the length of the intervals through
experimentations after implementing the relaxed schedule as part of the
skin triangulation algorithm.

\section{Analysis}
\label{sec4}

In this section, we derive lower bounds on the amount of time it takes
for an edge or triangle to change its size by more than some threshold value.
From these we will derive lower bounds on the time
it takes an element to pass through the entire size buffer.
We begin by studying the motion of a single point.

\paragraph{Traveling point.}
We recall that the speed of a point $u$ on the skin surface is
$\norm{\dot{u}} = 1/(2\norm{u})$,
assuming we write the patch that contains it in Standard Form.
The distance traveled by $u$ in a small time interval is therefore maximized if
it heads straight toward the origin, which for example happens if
$u$ lies on a shrinking sphere.
Starting the motion at point $u_0$, which is the point $u$
at time $t_0$, we get
\begin{eqnarray}
  \norm{u}  ~=~  \sqrt{ \norm{u_0}^2 - (t - t_0) } ,
  \label{eqn:norm}
\end{eqnarray}
for the point $u$ at time $t$.
This implies $t - t_0 = \norm{u_0}^2 - \norm{u}^2$,
so we see that $u$ reaches the origin at time $t = t_0 + \norm{u_0}^2$.
More generally, we reach the point $u_1 = (1-\theta) u_0$ between $u_0$
and the origin at time
$t_1  =  t_0 + \norm{u_0}^2 - \norm{u_1}^2 =  t_0 + (2 \theta - \theta^2) \norm{u_0}^2$.
Since the above analysis assumes the fastest way $u$ can possibly travel,
this implies that within an interval of duration $\Delta t = t_1 - t_0$,
the point $u_0$ cannot travel further than a distance
$\theta \varrho (u_0)$.
We use $\theta$ as a convenient intermediate quantity that gives us
indirect access to the important quantity, which is $\Delta t$.

Recall from the Curvature Variation Lemma of \cite{CDES01}
that the difference in length scale between two points is at most
the Euclidean distance.
If that distance is $\dist{u_0}{u_1} \leq \theta \varrho (u_0)$
then the length scale at $u_1$ is between $1-\theta$ and $1+\theta$
times the length scale at $u_0$.
It follows that if we travel for a duration
$\Delta t = (2 \theta - \theta^2) \varrho^2 (u_0)$,
we can change the length scale only by a factor
\begin{eqnarray}
  1 - \theta  ~\leq~  \frac{\varrho (u_1)}{\varrho (u_0)}
               ~<~  1 + \theta .
  \label{eqn:length_scale}
\end{eqnarray}
The lower bound is tight, and the upper bound cannot be reached because
the distance $\theta \varrho^2 (u_0)$ from $u_0$ can only be achieved
if the length scale shrinks.
We will also be interested in the integral of $1 / (2 \norm{u}^2)$,
which is again maximized if $u$ moves straight toward the origin:
\begin{eqnarray*}
 \int_{t_0}^{t_1} \frac{\diff t}{2 \norm{u}^2}
      ~&\leq&~  \int_{t_0}^{t_1} \frac{\diff t}{2 \norm{u_0}^2 - (2t - 2t_0)}    \\
      &=&~  (- \frac{1}{2}) \ln \frac{\norm{u_0}^2 - (t_1 - t_0)} {\norm{u_0}^2}  \\
      &=&~  \ln \frac{\norm{u_0}}{\norm{u_1}} .
\end{eqnarray*}
Denoting the above integral by $X$ and choosing
$t_1 - t_0 = (2 \theta - \theta^2) \norm{u_0}^2$, as before, we have
\begin{eqnarray}
  e^X  ~\leq~    \frac{\norm{u_0}}{\norm{u_1}}
        ~=~   \frac{\varrho(u_0)}{\varrho(u_1)}
        ~\leq~ \frac{1}{1-\theta} .
  \label{eqn:exponential}
\end{eqnarray}

\paragraph{Edge length variation.}
Consider two points $u$ and $v$ on the skin surface
during a time interval $[t_0, t_1]$.
We assume that both points follow their trajectories undisturbed
by any mesh maintenance operations.
Let $u_0$ and $u_1$ be the point $u$ at times $t_0$ and $t_1$
and, similarly, let $v_0$ and $v_1$ be the point $v$ at these two moments.
We prove that if the time interval is short relative to the
length scale of the points then the distance between them cannot
shrink or grow by much.
\begin{description}
 \item[{\sc Length Lemma.}]
  Let $\varrho_0 = \min \{ \varrho (u_0), \varrho (v_0) \}$ and
  $\Delta t = t_1 - t_0 = (2 \theta - \theta^2) \varrho_0^2$,
  for some $0 \leq \theta \leq 1$.
  Then
  \begin{eqnarray*}
   1 - \theta  ~\leq~  \frac{ \dist{u_1}{v_1} }{ \dist{u_0}{v_0} }
               ~<~  \frac{1}{1-\theta} .
  \end{eqnarray*}
\end{description}
\proof
 The derivative of the distance between points $u$ and $v$ with respect to time is
 \begin{eqnarray}
  \frac{\diff \dist{u}{v}}{\diff t}
    ~&=&~  \frac{\diff \dist{u}{v}}{\diff u} \frac{\diff u}{\diff t}
       + \frac{\diff \dist{u}{v}}{\diff v} \frac{\diff v}{\diff t} \nonumber \\
    ~&=&~  \frac{(u-v)^T}{\dist{u}{v}} (\dot{u} - \dot{v}) .
    \label{eqn:derivative}
 \end{eqnarray}
 For example if $u$ and $v$ lie on a common sphere patch then
 $\varrho = \varrho (u) = \varrho (v)$,
 $\dot{u} = {\pm u} / (2 \varrho^2)$ and $\dot{v} = {\pm v} / (2 \varrho^2)$,
 which implies
 \begin{eqnarray*}
  \frac{\diff \dist{u}{v}}{\diff t}
    ~=~  \pm \frac{(u-v)^T}{\dist{u}{v}} \frac{(u-v)}{2 \varrho^2}
   ~=~ \pm \frac{\dist{u}{v}}{2 \varrho^2} .
 \end{eqnarray*}
 We prove below that in the general case, the distance derivative stays
 between these two extremes:
 \begin{eqnarray}
  - \frac{\dist{u}{v}}{2 \varrho^2}
     ~\leq~  \frac{\diff \dist{u}{v}}{\diff t}
     ~\leq~ \frac{\dist{u}{v}}{2 \varrho^2} ,
  \label{eqn:length-bounds}
 \end{eqnarray}
 where $\varrho = \min \{ \varrho (u), \varrho (v) \}$.
 To get the final result from (\ref{eqn:derivative}),
 we divide by $\dist{u}{v}$, multiply by $\diff t$,
 and use $\diff \ln x = \diff x / x$ to get
 \begin{eqnarray*}
  - \frac{\diff t}{2 \varrho^2}  ~\leq~  \diff (\ln \dist{u}{v})
       ~\leq~  \frac{\diff t}{2 \varrho^2} .
 \end{eqnarray*}
 Next we integrate over $[t_0, t_1]$ and exponentiate to eliminate
 the natural logarithm:
 \begin{eqnarray*}
  {e^{-X}} ~\leq~  \frac{\dist{u_1}{v_1}}{\dist{u_0}{v_0}}
           ~\leq~  e^X .
 \end{eqnarray*}
 The claimed pair of inequalities follows from (\ref{eqn:exponential})
 and the observation that the upper bound for $X$ cannot be realized
 when the distance derivative is positive.
 To prove (\ref{eqn:length-bounds}) for general points $u$ and $v$,
 it suffices to show that the length of $\dot{u} - \dot{v}$
 is at most $\dist{u}{v} /(2 \varrho^2)$.
 We have seen that this is true if $u$ and $v$ belong to a common sphere patch.
 It is also true if $u$ and $v$ belong to a common hyperboloid patch
 because
 \begin{eqnarray*}
   \dist{\dot{u}}{\dot{v}}
    ~=~  \| {\frac{u'}{2 \varrho^2 (u)}} - {\frac{v'}{2 \varrho^2 (v)}} \|
    ~\leq~  \frac{\dist{u}{v}}{2 \varrho^2} ,
 \end{eqnarray*}
 where the primed and unprimed vectors are the same, except that they have a
 different sign in the third coordinate.
 We need a slightly more elaborate argument if $u$ and $v$ do not
 belong to the same mixed cell.
 We then reflect $v$ across the faces of mixed cells that intersect
 the edge $uv$.
 As described in Section \ref{sec2}, such a sequence of reflections
 does not affect the velocity vector.
 The distance between $u$ and the image $\overline{v}$ of $v$ under the composition
 of reflections is at most that between $u$ and $v$.
 Hence,
 \begin{eqnarray*}
   \dist{\dot{u}}{\dot{v}}  ~=~  \dist{\dot{u}}{\dot{\overline{v}}}
      ~\leq~  \norm{ \frac{u-v}{2 \varrho^2} } ,
 \end{eqnarray*}
 as required.
\eop

The lower bound in the Length Lemma is tight and realized
by points $u$ and $v$ on a common sphere patch.

\paragraph{Shrinking edge.}
Consider an edge $uv$, whose half-length at time $t_0$ is $R_0$.
As before, let $u_0$ and $v_0$ be the points $u$ and $v$ at
time $t_0$.
Let $\varrho_0 = \min \{ \varrho (u_0), \varrho (v_0) \}$.
We follow the two points during the time interval $[t_0, t_1]$,
whose duration is
$\Delta t = t_1 - t_0 = (2 \theta - \theta^2) \varrho_0^2$.
The Length Lemma implies that at time $t_1$,
the length of the edge satisfies
\begin{eqnarray}
  \frac{\dist{u_1}{v_1}}{\dist{u_0}{v_0}} 
    ~=~  \frac{R_1}{R_0}  ~\geq~  1 - \theta .
  \label{eqn:low_bnd_edge_shrink}
\end{eqnarray}
Our goal is to choose $\theta$ such that the edge at time $t_1$
is guaranteed to satisfy the Lower Size Bound for $Q = Q_1$.
Using $R_1 \geq (1-\theta) R_0$ from (\ref{eqn:low_bnd_edge_shrink}) and
$\varrho_1 < (1+\theta) \varrho_0$ from (\ref{eqn:length_scale}),
we note that $R_1 / \varrho_1 > C/Q_1$ is implied by
$(1-\theta) R_0 / (1+\theta) \geq C \varrho_0 / Q_1$.
In other words,
\begin{eqnarray}
  \theta ~=~ \frac{R_0 Q_1 - C \varrho_0}{R_0 Q_1 + C \varrho_0}
  \label{eqn:theta}
\end{eqnarray}
is sufficiently small.
The corresponding time interval during which we can be sure
that the edge $uv$ does not become unacceptably short has duration
$\Delta t = (2 \theta -\theta^2) \varrho_0^2$.
To get a better feeling for what these results mean, let us
write the half-length of $u_0 v_0$ as a multiple of the lower bound in [L]
for $Q = Q_0$:  $R_0 = A C \varrho_0 / Q_0$ with $A > 1.0$.
We then get $\theta = ({A Q_1 - Q_0})/({A Q_1 + Q_0})$ and
$\Delta t$ from $\theta$ as before.
Table \ref{tab:edge_constants} shows the values of $\theta$
and $\Delta t$ for a few values of $A$.
\begin{table}[hbt]
 \centering
 \caption{For edges, the values of $\theta$ and $\Delta t$ 
          for $Q_0 = 1.6$, $Q_1 = 2.3$ 
          and a few typical values of $A$}
 \label{tab:edge_constants}
 \begin{tabular}{|r||r|r|} \hline
   \multicolumn{1}{|c||}{$A$} 
         & \multicolumn{1}{c|}{$\theta$}
                            &  \multicolumn{1}{c|}{$\Delta t/\varrho_0^2$}  \\ \hline \hline
   $1.0$ &  $0.179\ldots$   &           $0.326\ldots$  \\ 
   $1.5$ &  $0.366\ldots$   &           $0.598\ldots$  \\
   $2.0$ &  $0.483\ldots$   &           $0.733\ldots$  \\
   $2.5$ &  $0.564\ldots$   &           $0.810\ldots$  \\
   $3.0$ &  $0.623\ldots$   &           $0.858\ldots$  \\
   $3.5$ &  $0.668\ldots$   &           $0.890\ldots$  \\
   $4.0$ &  $0.703\ldots$   &           $0.912\ldots$  \\ \hline
 \end{tabular}
\end{table}

\paragraph{Height variation.}
Consider a triangle $uvw$ during a time interval $[t_0, t_1]$.
We assume that all three points follow their trajectories undisturbed
by any mesh maintenance operations.
Each vertex has a distance to the line spanned by the other two vertices,
and the \emph{height} $H$ of $uvw$ is the smallest of the three distances.
If $uv$ is the longest edge then $H = \dist{w}{w'}$,
where $w'$ is the orthogonal projection of $w$ onto $uv$.
We prove if the time interval is short relative to the length
scale at the points then the height cannot shrink or grow by much.
To state the claim we use indices 0 and 1 for points and heights
at times $t_0$ and $t_1$.
\begin{description}
 \item[{\sc Height Lemma.}]
  Let $\varrho_0 = \min \{ \varrho (u_0), \varrho (v_0), \varrho (w_0) \}$ and
  $\Delta t = t_1 - t_0 = (2 \theta - \theta^2) \varrho_0^2$,
  for some $0 \leq \theta \leq 1$.
  Then
  \begin{eqnarray*}
    1 - \theta  ~\leq~  \frac{H_1}{H_0} ~<~ \frac{1}{1-\theta} .
  \end{eqnarray*}
\end{description}
\proof
 We prove that (\ref{eqn:length-bounds}) is also true if we
 substitute the height $H$ for the length of the edge $uv$:
 \begin{eqnarray}
  - \frac{H}{2 \varrho^2}  ~\leq~  \frac{\diff H}{\diff t}
       ~\leq~  \frac{H}{2 \varrho^2} ,
  \label{eqn:height-bounds}
 \end{eqnarray}
 where $\varrho = \min \{ \varrho (u), \varrho (v), \varrho (w) \}$.
 The claimed pair of inequalities follows as explained in the
 proof of the Length Lemma.
 To see (\ref{eqn:height-bounds}) note first that the height of the
 triangle is always determined by a vertex and a point on the opposite
 edge, eg.\ $H = \dist{w}{w'}$.
 Let $w' = (1-\lambda) u + \lambda v$.
 If $u$ and $v$ belong to the same mixed cell then
 $\nabla g_{w'} = (1-\lambda) \nabla g_u + \lambda \nabla g_v$
 because the gradient varies linearly.
 Along a moving line segment $uv$ the velocity vectors
 vary linearly, hence $\dot{w}' = (1-\lambda) \dot{u} + \lambda \dot{v}$.
 Since the gradients and the velocity vectors at $u$ and $v$
 point in the same directions, they do the same at $w'$.
 The length of the velocity vector at $w'$ is at most that of the
 longer velocity vector at $u$ and $v$.
 If $w$ belongs to the same mixed cell as $w'$, this implies
 \begin{eqnarray*}
  \dist{\dot{w}}{\dot{w}'}  ~\leq~  \frac{\norm{w-w'}}{2 \varrho^2}
                            ~=~   \frac{H}{2 \varrho^2} ,
 \end{eqnarray*}
 from which (\ref{eqn:height-bounds}) follows.
 If $u$, $v$ and $w$ do not belong to the same mixed cell then we
 perform reflections, as in the proof of the Length Lemma,
 and get (\ref{eqn:height-bounds}) because reflections do
 not affect velocity vectors.
\eop

In the following, we only need the lower bound in the Height Lemma,
which is tight and is realized points $u$, $v$ and $w$ on a common
sphere patch.

\begin{figure}[hbt]
  \vspace*{0.1in}
  \centering
  \centerline{\includegraphics[height=1.8in]{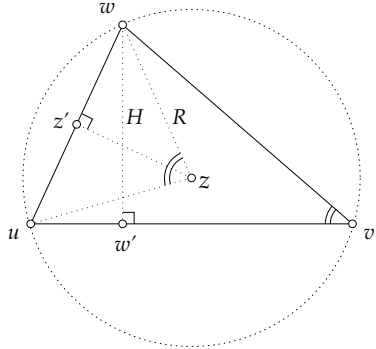}}
  \caption{The triangle $uvw$ is similar to $wz'z$, which implies a relation
           between the height $H$ and the circumradius $R$}
  \label{fig:radius_height}
\end{figure}

\paragraph{Expanding triangle.}
We use both the Length Lemma and the Height Lemma to derive a lower bound
on the length of time during which a triangle that initially satisfied the
Upper Size Bound [U] for $Q = Q_0$ is guaranteed to satisfy the same for $Q = Q_1$.
We begin by establishing a relation between the circumradius $R = R_{uvw}$
of a triangle $uvw$ and its height and edge lengths.
Referring to Figure \ref{fig:radius_height}, we let $z$ denote the center of
the circumcircle.
Assuming $uv$ is the longest of the three edges, the height is
$H = \dist{w}{w'}$ and $v$ and $z$ lie on the same side of the line
passing through $u$ and $w$.
Let $z'$ be the midpoint of $uw$ and note that the angle at $z$
is twice that at $v$:
$\angle uzw = 2 \angle z'zw = 2 \angle uvw$.
This implies that the triangles $ww'v$ and $wz'z$ are similar,
and therefore $\dist{z'}{w} / R = H / \dist{v}{w}$.
It follows that the circumradius of $uvw$ is
\begin{eqnarray*}
  R  ~=~  \frac{\dist{u}{w} \ \dist{v}{w}}{2 H} .
\end{eqnarray*}
There are three ways to write twice the area as the product of an edge length
and the distance of the third vertex from the line of that edge:
$\dist{u}{v} \ H = \dist{u}{w} \ \dist{v}{v'} = \dist{v}{w} \ \dist{u}{u'}$.
Hence, the circumradius is also
\begin{eqnarray*}
  R  ~=~  \frac{\dist{u}{v} \ \dist{u}{w}}{2 \dist{u}{u'}}
    ~=~ \frac{\dist{u}{v} \ \dist{v}{w}}{2 \dist{v}{v'}} .
\end{eqnarray*}
For the remainder of this section, we use indices 0 and 1 for points, heights
and radii at times $t_0$ and $t_1$.
The above equations for the circumradius imply
\begin{eqnarray*}
  \frac{R_1}{R_0}  ~=~
     \frac{\dist{u_1}{w_1}}{\dist{u_0}{w_0}}
     \frac{\dist{v_1}{w_1}}{\dist{v_0}{w_0}}
     \frac{\dist{w_0}{w'_0}}{\dist{w_1}{w'_1}} .
\end{eqnarray*}
Assuming $H_0 = \dist{w_0}{w_0'}$ is the height at time $t_0$,
we have $H_1 \leq \dist{w_1}{w_1'}$ at time $t_1$.
We can therefore use the Length Lemma to bound the first two ratios
and the Height Lemma to bound the third to get
\begin{eqnarray}
  \frac{R_1}{R_0}  ~<~  \frac{1}{(1-\theta)^3} .
  \label{eqn:radius-bound}
\end{eqnarray}
We now choose $\theta$ such that a triangle that satisfies [U] for $Q = Q_0$
at time $t_0$ is guaranteed to satisfy [U] for $Q = Q_{1}$ at time $t_1$.
Using $R_1 < R_0 / (1-\theta)^3$ from (\ref{eqn:radius-bound}) and
$(1-\theta) \varrho_0 \leq \varrho_1$ from (\ref{eqn:length_scale}), we note
that $R_1 / \varrho_1 < C Q_1$ is implied by
$R_0 / (1-\theta)^4 \leq C Q_1 \varrho_0$.
In other words,
\begin{eqnarray}
  \theta  ~=~  1 - \sqrt[4]{R_0 / (C Q_1 \varrho_0)}
  \label{eqn:triangle-theta}
\end{eqnarray} 
is sufficiently small.
It is convenient to write the circumradius of the triangle $u_0 v_0 w_0$
as a fraction of the upper bound in [U]:
$R_0 = C Q_0 \varrho_0 / A$ with $A > 1.0$.
Then, $\theta = 1 - \sqrt[4]{Q_0 / (A Q_1)}$.
Table \ref{tab:triangle_constants} shows the values of $\theta$
and $\Delta t$ for a few values of $A$.
\begin{table}[hbt]
 \centering
 \caption{For triangles, the values of $\theta$ and $\Delta t$ for
         $Q_0 = 1.6$, $Q_1 = 2.3$ and a few typical values of $A$}
 \label{tab:triangle_constants}
 \begin{tabular}{|r||r|r|} \hline
   \multicolumn{1}{|c||}{$A$} 
         & \multicolumn{1}{c|}{$\theta$}
                            &  \multicolumn{1}{c|}{$\Delta t/\varrho_0^2$}  \\ \hline \hline
   $1.0$ &  $0.086\ldots$   &           $0.165\ldots$  \\ 
   $1.5$ &  $0.174\ldots$   &           $0.319\ldots$  \\
   $2.0$ &  $0.232\ldots$   &           $0.410\ldots$  \\
   $2.5$ &  $0.273\ldots$   &           $0.472\ldots$  \\
   $3.0$ &  $0.306\ldots$   &           $0.518\ldots$  \\
   $3.5$ &  $0.332\ldots$   &           $0.554\ldots$  \\
   $4.0$ &  $0.354\ldots$   &           $0.583\ldots$  \\ \hline
 \end{tabular}
\end{table}

\section{Discussion}
\label{sec5}

The main contribution of this paper is the introduction of relaxed scheduling
as a paradigm for maintaining moving or deforming data,
and the demonstrations of its applicability to scheduling edge contractions
and vertex insertions maintaining skin surfaces.

\paragraph{Algorithm design.}
We view the dynamic skin triangulation algorithm, of which relaxed scheduling
is now a part, as an interesting exercise in rational algorithm design.
What are the limits for proving meshing algorithms correct?
This design exercise gives us a glimpse on how complicated meshing problems can be.
Perhaps more importantly, it illustrates what it might take to prove other
meshing algorithms correct.
We especially highlight the role of constant parameters in the algorithm
and how they control the algorithm as well as the constructed mesh.
In our example, the important parameters are $C$,
which controls how closely the mesh approximates the surface,
and $Q$, which controls the quality of the mesh.
The effort of proving the various pieces of the algorithm correct
has lead to inequalities for these parameters.
In other words, we have identified a feasible region which is necessary
for our proofs and sufficient for the correctness of the algorithm.
The detailed knowledge of this feasible region has inspired the idea
of relaxed scheduling, and it was necessary to formulate it in detail
and to prove its correctness.
Many meshing algorithms are based on parameters that are fine-tuned
in the experimental phase of software design.
We suggest that in the absence of detailed knowledge of limits,
fine-tuning is a necessary activity that gropes for a place in the
feasible region where correctness is implied.
Of course, it might happen that this region is empty, but this is
usually difficult to determine.

\paragraph{Future work.}
It is not our intention to criticize work in mesh generation for the
lack of correctness proofs.
Indeed, it would be more appropriate to criticize our own work for the
lack of generality.
Although we laid out a complete algorithm for maintaining the mesh
of a deforming surface, we are a far cry from being able to prove its
correctness for any surface other than the skin surface
introduced in \cite{Ede99}.
We have also not been able to extend the algorithm beyond the
deformations implied by growing the spheres that define the surface.
For example, it would be desirable to maintain the mesh for deformations
used for morphing as described in \cite{CEF01}.
Generalizing the algorithm to include this application and proving
it correct may be within reach.

Another worthwhile task is the implementation of relaxed scheduling
as part of the dynamic skin algorithm.
Are our lower bounds for the necessary $\Delta t$ sufficient to eliminate
edge contractions and vertex insertions as a bottleneck of the algorithm?
Can these lower bounds be improved in any significant manner?
Can we improve the performance by fine-tuning the parameters, in particular
$Q_0$ and $Q_1$, while staying within the proved feasible region?

\section*{Acknowledgments}
The authors thank Robert Bryant for helpful discussions concerning the
proof of the Length Lemma.



\section*{Appendix}

\vspace*{-.2in}

\begin{table}[hbt]
  \centering
  \caption{Notation for important geometric concepts,
           functions, variables, and constants}
  \label{tab:Notation}
  \begin{tabular}{ll}
    $\wdist{i} : \Rspace^3 \rightarrow \Rspace$
                         &  weighted (square) distance function      \\
    $S_i = (z_i, r_i)$   &  zero-set of $\wdist{i}$;                 
                          sphere with center $z_i$ and radius $r_i$\\
    $\Fset$              &  convex hull of spheres $S_i$             \\
    $F = \env{\sqrt{\Fset}}$ &  skin surface                         \\
    $\kappa, \varrho = 1/\kappa$
                         &  maximum curvature, length scale          \\
    $Q_0 \leq Q \leq Q_1$&  constant controlling quality             \\
    $\Eps, C, h, \ell, m$&  additional constants                     \\
                         &                                           \\
    $g: \Rspace^3 \rightarrow \Rspace$
                         &  point-wise min of the $2f-f(z)$          \\
    $F(t) = g^{-1}(t)$   &  skin surface at time $t$                 \\
    $t, \theta$          &  time parameter, relative travel distance    \\
    $[t_0, t_1]$         &  time interval                            \\
    $\Delta t = t_1-t_0$ &  duration                                 \\
    $u, u', \overline{u}$ & point, projection, reflection            \\
    $\nabla g_u, \dot{u}$ & gradient, velocity vector                \\
    $uv, uvw, H, R$      &  edge, triangle, height, radius
  \end{tabular}
\end{table}

\end{document}